\documentclass[12pt,preprint]{aastex}

\def\arcsec{\ifmmode '' \else $''$\fi}
\def\arcmin{\ifmmode ' \else $'$\fi}
\def\arcsecpoint{\ifmmode ''\!. \else $''\!.$\fi}
\def\arcminpoint{\ifmmode '\!. \else $'\!.$\fi}
\def\cc{\ifmmode {\rm cm}^{-3} \else cm$^{-3}$\fi}
\def\cl{\ifmmode {\rm cm}^{-2} \else cm$^{-2}$\fi}
\def\micron{\ifmmode \mu{\rm m} \else $\mu$m\fi}
\def\kms{\ifmmode {\rm km\,s}^{-1} \else km\,s$^{-1}$\fi}
\def\Hubble{\ifmmode {\rm km\,s}^{-1}\,{\rm Mpc}^{-1}
        \else km\,s$^{-1}$\,Mpc$^{-1}$\fi}
\def\ergsec{\ifmmode {\rm ergs\;s}^{-1} \else ergs s$^{-1}$\fi}
\def\ergscm{\ifmmode {\rm ergs\,s}^{-1}\,{\rm cm}^{-2}
          \else ergs\,s$^{-1}$\,cm$^{-2}$\fi}
\def\ergscmA{\ifmmode {\rm ergs\,s}^{-1}\,{\rm cm}^{-2}\,{\rm \AA}^{-1}
          \else ergs\,s$^{-1}$\,cm$^{-2}$\,\AA$^{-1}$\fi}
\def\ergscmHz{\ifmmode {\rm ergs\,s}^{-1}\,{\rm cm}^{-2}\,{\rm Hz}^{-1}
          \else ergs\,s$^{-1}$\,cm$^{-2}$\,Hz$^{-1}$\fi}
\def\Msun{\ifmmode M_{\odot} \else $M_{\odot}$\fi}
\def\Lsun{\ifmmode L_{\odot} \else $L_{\odot}$\fi}
\def\qo{\ifmmode q_{0} \else $q_{0}$\fi}
\def\Ho{\ifmmode H_{0} \else $H_{0}$\fi}
\def\h{H {\rm I}}

\def\ciii{C\,{\rm III}}
\def\civ{C\,{\rm IV}}
\def\nv{N\,{\rm V}}
\def\oiii{O\,{\rm III}}
\def\ari{Ar\,{\rm I}}

\newcommand{\oi}{O~{\rm I}}
\newcommand{\ovi}{O~{\rm VI}}
\newcommand{\ovii}{O~{\rm VII}}
\newcommand{\oviii}{O~{\rm VIII}}
\newcommand{\heii}{He~{\rm II}}

\newcommand{\niii}{N~{\rm III}}

\begin{document}

\title{Far Ultraviolet Spectroscopic Explorer Observations of \\
the Seyfert 1.5 Galaxy NGC 5548 in a Low State} 

\author{
M. S. Brotherton\altaffilmark{1},
R. F. Green\altaffilmark{1},
G. A. Kriss\altaffilmark{2,3},
W. Oegerle\altaffilmark{4},
M. E. Kaiser\altaffilmark{3},
W. Zheng\altaffilmark{3},
J.~B.~Hutchings\altaffilmark{5}
}
\altaffiltext{1}{Kitt Peak National Observatory,
        National Optical Astronomy Observatories, P.O. Box 26732,
        950 North Cherry Ave., Tucson, AZ, 85726-6732; mbrother, green@noao.edu}
\altaffiltext{2}{Space Telescope Science Institute,
        3700 San Martin Drive, Baltimore, MD 21218}
\altaffiltext{3}{Center for Astrophysical Sciences, Department of Physics and
        Astronomy, The Johns Hopkins University, Baltimore, MD 21218--2686}
\altaffiltext{4}{Laboratory for Astronomy and Solar Physics, Code 681,
        NASA/Goddard Space Flight Center, Greenbelt, MD 20771}
\altaffiltext{5}{Herzberg Institute of Astrophysics, National Research Council 
	of Canada, Victoria, BC, V9E 2E7, Canada; john.hutchings@hia.nrc.ca}

\begin{abstract}
We present far-ultraviolet spectra of the Seyfert 1.5 galaxy NGC~5548 obtained 
in 2000 June with the Far Ultraviolet Spectroscopic Explorer (FUSE).
Our data span the observed wavelength range 915--1185 \AA\ at a resolution
of $\sim20~\kms$.  The spectrum shows a weak continuum and emission from
\ovi~$\lambda\lambda$1032,1038, \ciii~$\lambda$977, and 
\heii~$\lambda$1085.  
The FUSE data were obtained when the AGN was in
a low state, which has revealed strong, narrow \ovi\ emission lines.
We also resolve intrinsic, associated absorption lines of \ovi\ 
and the Lyman series.  Several distinct kinematic components are present, 
spanning a velocity range of $\sim 0~\rm to\ -1300~\kms$ relative to systemic,
with kinematic structure similar to that seen in previous observations of 
longer wavelength ultraviolet (UV) lines.  We explore the relationships 
between the far-UV absorbers and those seen previously in the UV and X-rays.  
We find that the high-velocity UV
absorption component is consistent with being low-ionization, 
contrary to some previous claims, and is consistent with its non-detection 
in high-resolution X-ray spectra.
The intermediate velocity absorbers, at $-300~\rm to\ -400~\kms$, show
\h\ and \ovi\ column densities consistent with having contributions from
both a high-ionization X-ray absorber and a low-ionization UV absorber.
No single far-UV absorbing component can be solely identified with 
the X-ray absorber.

\end{abstract}


\section{Introduction}

NGC 5548 is among the most studied of Seyfert galaxies.  
As a bright, nearby (z=0.017175)
variable AGN, NGC 5548 in particular has been the target of intensive 
monitoring campaigns in the optical (e.g., Peterson et al. 1999), the
ultraviolet (Clavel et al. 1991; Korista et al. 1995), and the
extreme ultraviolet (Marshall et al. 1997).  The time-delayed
response of emission lines to continuum changes indicates the size-scale
of the emitting regions.  In the case of NGC 5548, the Doppler widths
of the lines and their respective size scales are consistent with 
Keplerian motion around a black hole of mass $5.9 \times 10^7 M_{\sun}$
(Peterson \& Wandel 2000).
  
NGC 5548 is also noteworthy as an AGN with intrinsic ultraviolet (UV) and 
X-ray absorption characteristic of a high-velocity outflow.  
About half of Seyfert galaxies display complex 
narrow intrinsic absorption of high-ionization species with 
transitions in the UV (Crenshaw et al. 1999).  Moreover, there 
appears to be a one-to-one correspondence between the
presence of this UV absorption and warm X-ray absorption.
Warm absorbers appear to be highly ionized (typically U = 0.1 to 10,
where U is the ratio of the number density of hydrogen-ionizing photons to 
the hydrogen number density) with total column densities in the range 
10$^{21-23}$ cm$^{-2}$ (George et al. 1998). 

The correspondence between the presence of UV absorption and of
X-ray absorption suggests a direct link between the two despite
the ionization difference.  Mathur et al. (1995) proposed a single
zone model for a combined UV-X-ray absorber in NGC 5548.
High-resolution UV spectroscopy shows that the UV absorber is resolved into
distinct kinematic components with outflow velocities ranging to 
greater than 1100 \kms\ (Mathur et al. 1999; Crenshaw \& Kraemer 1999),
and with a range of ionization states and column densities.

The new generation of X-ray telescopes (Chandra, XMM) now provide the 
combination of sensitivity and spectral resolution to detect and
measure the kinematic properties of X-ray absorption lines in AGNs.
Kaastra et al. (2000) presented the first high-resolution X-ray spectrum of
a Seyfert galaxy, in fact, NGC 5548.  The Chandra-LETGS spectrum revealed
strong narrow absorption lines from highly ionized species, blueshifted
by several hundred \kms.  

The Far Ultraviolet Spectroscopic Explorer (FUSE) (Moos et al. 2000) 
provides spectral coverage 
of shorter wavelengths than possible with the Hubble Space Telescope (HST).
Of special interest in this wavelength range for investigations of the 
intrinsic UV-X-ray absorbers is the \ovi~$\lambda\lambda$1032,1038
resonance doublet, as
\ovi\ is of intermediate ionization.  In the warm absorber in Markarian 509, 
for instance, FUSE resolves seven kinematic components of intrinsic \ovi\ 
absorption, one of which -- and only one of which -- can be identified with
the X-ray absorber (Kriss et al. 2000).  
Kriss (2000) reviews this result along with other FUSE observations of AGNs.

In this paper we present FUSE observations of NGC 5548 (\S~2), focusing in 
particular on characterizing the intrinsic \ovi\ and Lyman $\beta$ absorption
components (\S~3).   We interpret the spectral features with the assistance
of photoionization models (\S~4), and discuss our findings together with
the previous results from the UV and X-ray to achieve an improved understanding
of the ionized outflow of NGC 5548 (\S~5).  

\section{Observations}
FUSE (Moos et al. 2000; Sahnow et al. 2000) contains four ``channels'' which 
are in effect four independent telescope/grating spectrographs, two of which 
employ optics coated with LiF for high efficiency from $\sim990$--1187 \AA\ 
and two of which employ optics coated with SiC for high efficiency at shorter 
wavelengths.  Two two-dimensional photon-counting microchannel-plate detectors 
record spectra from the four channels.

NGC 5548 was observed on 2000 June 7 through the $30\arcsec \times 30\arcsec$
low-resolution aperture with a total exposure time of 25 ksec.
Data were recorded in photon-address mode, which provides a time-tagged list
of event positions in the down-linked data stream.
All channels were aligned, resulting in spectra covering the 
whole wavelength range from 905 to 1187 \AA.  

To calibrate the data, we use a modified version of the standard FUSE
pipeline.  Our processing restricts the pulse heights for acceptable data
to channels 4--16 to reduce the background.  We also scale the background
level to match the levels seen in blank regions of the detector.
Because FUSE uses the LiF1 channel for fine guidance, this channel has
the best photometric accuracy and the best absolute wavelength scale.
We therefore cross-correlated the other three channels with LiF1
to obtain adjustments to the zero points of their wavelength scales,
and we scaled their fluxes to match the level seen in LiF1.
To obtain the best signal-to-noise ratio (S/N) in our final calibrated
spectrum, we then re-binned the data from all the detector segments
onto a uniform wavelength scale using 0.05 \AA\ bins.  Approximately
7 pixels from each detector segment contributes to each bin. This results
in an effective spectral resolution of $\sim20~\kms$.  We estimate that
our fluxes are accurate to $\sim$10\%, and that the wavelength scale has
an absolute accuracy of $\sim15~\kms$.

Figure 1 displays the full FUSE spectrum.
We note the presence of emission lines of \ovi~$\lambda\lambda$1032,1038, 
\ciii~$\lambda$977, \niii~$\lambda$990, and \heii~$\lambda$1085.
We do not see an intrinsic Lyman edge.
Galactic molecular hydrogen appears present but weak along this sight line, 
and a number of atomic lines from the Galactic ISM are present as well 
(including \ovi~$\lambda\lambda$1032,1038).  A Galactic Lyman edge is also seen.

NGC 5548 was in a low state when observed by FUSE.  Coincidently, this low
state appears to match the flux level of NGC 5548 in 1991 when first observed 
with the Faint Object Spectrograph (FOS) on the Hubble Space Telescope (HST)
(Crenshaw, Wu, \& Bogges 1993).  This level is a factor of several lower
than when observed  in recent years by, for instance, the Hopkins Ultraviolet
Telescope (Kriss et al. 1997).  
Figure 2 compares these spectra.  In the low state
the continuum and broad lines weaken and the less time-variable narrow lines
become more prominent.

\section{Spectral Measurements}

Given the low state of NGC 5548, the signal-to-noise ratio (S/N) of the 
spectrum is not as high as initially anticipated.  
For $\lambda < 1000$ \AA\, the S/N per 0.05 \AA\ pixel is only about 3;
we therefore cannot make accurate measurements of \ciii~$\lambda$977 and
the higher order Lyman lines, although some degree of absorption is 
probably present (contaminating Galactic \oi~$\lambda$989 certainly is 
present).  At longer wavelengths in the Ly $\beta$ and \ovi\ region, 
the S/N is about 10 in the continuum and higher in the emission lines.
This is large enough for detailed work at the full resolution of FUSE.

We have made measurements of the Lyman $\beta$/\ovi\ region using our 
0.05 \AA\ combined-channel spectrum.  We have used SPECFIT (Kriss 1994) to fit 
the spectral region.  The continuum is modeled as a power-law plus
Galactic extinction (A$_B$ = 0.088 mag, Schlegel et al. 1998) using the
reddening curve of Cardelli, Clayton, \& Mathis (1989) with $R_V = 3.1$. 
The \ovi\ emission lines are fit using Gaussian components.
Since inflections are present in the emission profiles, these separate 
components may have physical significance.
\ovi~$\lambda\lambda$1032,1038 is a doublet with an optically thin line ratio 
of 2:1, but in dense gas with saturation the ratio may approach 1:1.  
While the physical conditions typically associated 
with the narrow-line region (NLR) should only show the 2:1 
\ovi~$\lambda\lambda$1032,1038 doublet emission ratio, the ratio is 1:1
in the low-state FUSE spectrum of NGC 3516 (Hutchings et al. 2001).
Given this empirical result and the possibility of narrow
\ovi~$\lambda\lambda$1032,1038 emitting gas with atypical conditions,
our initial models begin with 2:1 doublet ratios but are allowed to vary
between 2:1 and 1:1.

We approximate the intrinsic absorption as several components with
Gaussian distributions in optical depth that are permitted to partially
cover the background source.  The optical depth ratio of the doublets is 
constrained to be 2:1 at all velocities.  The initial parameters for the 
Gaussians (e.g., FWHM, redshift) were selected to match those reported for 
UV lines (components 1--6) by Crenshaw \& Kraemer (1999), but allowed to 
freely vary to account for source variability, ionization differences, and 
data quality.  We also added two new components, a high-velocity component 
``0.5'' and a component ``2.5'', which significantly improved the fits.  
The absorption lines of the \ovi\ doublets and Ly $\beta$ have been fit
simultaneously and constrained to have the same covering
fraction\footnote{The partial covering of absorption components
in principle can vary between red and blue doublets since the ratios of
different emission components, with potentially different individual 
coverings, can differ.  In instances like this one, permitting this extra 
freedom in the fitting, however, compromises the consistency of the results.}.

Column densities for a particular ionized species are obtained from
measurements of optical depth according to the equation:
\begin{equation}
N_{ion} = \frac{3.7679\times10^{14}{\rm cm}^{-2}}{\lambda f} \times \int \tau (v) dv\ \kms
\end{equation}
\noindent
where $\lambda$\ is the transition wavelength and $f$\ is the pertinent
oscillator strength.  The units on columns discussed below will always
be cm$^{-2}$, which will be implicitly understood for the case of logarithmic
columns.  For \ovi~$\lambda$1032, log $\lambda f = 2.137$\ and
for Lyman $\beta$ log $\lambda f = 1.909$\ (Morton 1991), where $\lambda$
is in \AA.
For the case of absorption lines that have Gaussians in optical depth:
\begin{equation}
\int \tau (v) dv = 1.0645  \times {\rm FWHM} \times \tau_0
\end{equation}
\noindent
where $\tau_0$ is the optical depth at zero velocity,
and the FWHM is in \kms.

Contaminating Galactic lines are included in the models.
We note in particular \oi~$\lambda$1039 and \ari~$\lambda$1048 fall at the same
wavelengths as redshifted intrinsic absorption troughs (\ari~$\lambda$1048 is
constrained by simultaneously fitting \ari~$\lambda$1066).  The atomic
transitions are constrained to have the same line widths.  Several strong
Galactic molecular hydrogen lines absorb some of the blue peak of \ovi\ 
emission.

\subsection{Specific Models}

There are two primary classes of models that yield plausible fits 
to the spectrum.  The first class, and the one we favor, features 
narrow emission lines with FWHM $\sim 400~\rm km~s^{-1}$ that are uncovered 
by absorption components.  These emission lines may be associated with
a traditional, extended, low-density NLR.  A small
degree of skewness in the form of a blue tail is allowed by the model and
consistent with NLR profiles (e.g., Brotherton 1996).  The weak broad emission
is fit using the sum of two Gaussians.  
These emission lines are described in Table 1 and shown in green in Figure 3.  
The redshift of the narrow emission lines 
in this case is 0.01702, only 50 \kms\ blueshifted from the 
21 cm redshift $z$ = 0.017175, and clearly larger than the ``emission-line 
redshift'' of $z$ = 0.01676 given in the NASA/IPAC Extragalactic Database (NED).
Table 2 gives the parameters for the absorption components in our best
fitting model.  

The second class of models involves a significant extrapolation of the 
narrow emission components, in which we assume that the NLR is at least
partially covered by the absorbers.  The motivations for this type of model 
are that the
high-ionization NLR is expected to be compact (Kraemer et al. 1998) and hence 
more easily covered, that the narrow-line component of the UV lines seen in the 
low-state FOS spectrum have FWHM $\sim$ 1000 \kms\ (Goad \& Koratkar 1998),
and that low-state narrow \ovi\ emission appears to arise at small radii
at the base of an accelerating wind in NGC 3516 (Hutchings et al. 2001).
Our best covered NLR model has a narrow emission component with FWHM 
$\sim 650~\rm km~s^{-1}$, larger than that of the uncovered case and
more similar to the Goad \& Koratkar (1998) results.
Figure 4 shows this model, and Tables 1 and 2 give the model parameters.
Table 1 reveals that the narrow \ovi\ doublet ratio is an optically thin 2:1,
unlike the case of NGC 3516 where it is nearly 1:1 (Hutchings et al. 2001),
weakening the similarity between the two AGNs in the low state.

While we favor the uncovered NLR model as it is less of an extrapolation
and yields a better $\chi^2$, an examination of Table 2 reveals that 
both models give very similar results for the absorption lines.  
Despite the possible uncertainty about the FWHM and covering of the narrow 
emission lines, the inferred absorption line column densities and ratios appear 
rather robust.

\subsection{Continuum}

We also have made independent fits to measure the continuum slope, expanding
the fitting range beyond the immediate vicinity of \ovi\ and Lyman $\beta$,
and excluding portions of spectra with clear emission or
absorption present.  Assuming only a power law continuum shape 
and Galactic reddening, the power-law index obtained is 
$\alpha$ = $-$1.23$\pm$0.11, where $F_{\nu} \propto \nu^{\alpha}$. 

\section{Intrinsic Absorption in NGC 5548}

We interpret the absorbing column densities in Table 2 with the assistance of 
photoionization models.
We have employed the latest version of Cloudy (C94; Ferland 2001).  We
assume solar abundances and the continuum shape of Dumont et al. (1998).
We characterize the state of the gas in terms of the dimensionless 
ionization parameter U, which is the ratio of the number density of 
hydrogen-ionizing photons to the number density of hydrogen.
Within our FUSE spectrum we can measure simultaneous column densities only
for neutral hydrogen and for \ovi.  The physical conditions of an absorber 
are not uniquely determined with only \h\ and \ovi\ column densities. 
In general there will be two solutions, one with a low ionization parameter 
and low total column density, and another with a high ionization parameter 
and higher column density.  For gas in both very low and very high
ionization states, there is no \ovi\ as all the oxygen is in lower
or higher ionization states, respectively.  As a function of increasing
ionization parameter, the ratio of \h\ to \ovi\ column densities starts at 
zero, rises and peaks ($\sim$10 near U$\sim$0, for our models), then falls 
again.  The Cloudy command OPTIMIZE has been used to find solutions without 
creating large grids in parameter space.  Table 3 gives both solutions for the 
column densities measured for each component in our preferred uncovered NLR fit.

Although the formal errors on many component parameters are small, there is a
certain degree of arbitrariness in the models resulting from the
blending of the fitted components (especially in conjunction with partial
covering).  With the addition of two components and freely varying velocity
widths, and possible variability, a component-to-component comparison with UV
absorbers seen by HST (e.g., Crenshaw \& Kraemer 1999; Mathur et al.  1999) 
remains imperfect.  Figure 5 shows the velocities of the six HST components 
on a blow up of the blue \ovi\ doublet; it appears that the FUSE spectrum
is redshifted some 100 \kms\ relative to the STIS spectrum.
An examination of Galactic lines in the STIS spectrum using on-the-fly 
calibration suggests that 50 \kms\ of the shift may arise there.
Potential variability and low spectral resolution 
in the X-rays also limit comparison with Chandra spectra (Kaastra et al. 2000).
Still, there is much to be learned from such comparisons assisted by 
our photoionization models.  We do so below.

\subsection{Components 0.5 and 1}
We first consider high-velocity components ($-1000$ to $-1300$ \kms).
The blue \ovi\ absorption at these velocities is beyond any uncertainties
in the \ovi\ narrow emission lines.  Component 1 is also of special interest 
because Crenshaw \& Kraemer (1999) claim, based on the column
density ratio of N (\nv)/N (\civ) seen in GHRS HST spectra, that this component 
has a high ionization parameter and large column density and could be a 
manifestation of the X-ray absorber.  

Crenshaw \& Kraemer (1999) started with N (\nv)/N (\civ) = $18\pm7$, based on
non-simultaneous GHRS spectra taken six months apart of the \nv\ and \civ\
spectral regions.  Their preferred physical conditions for a model to 
account for this single column density ratio are log U = 0.38, and effective 
hydrogen column density of log N$_H$ = 21.8.  They investigated varying
U or N$_H$\ in order to explain their later epoch STIS spectrum, for which
N (\nv) was lower, N (\h) was known, and N (\civ) had an upper limit.
These resulted in similarly high U, high N$_H$ conditions for the component 1
absorber.

Table 3 shows our pair of solutions for component 1 (ignoring the weaker and 
more uncertain component 0.5):
1) log U = 0.26, total hydrogen column log N$_H$ = 21.1;
2) log U = $-$1.71, total hydrogen column log N$_H$ = 18.4.
From Crenshaw \& Kraemer (1999), we might be led to assume our high-ionization 
solution represents the conditions in component 1.  The HST spectra were taken 
when NGC 5548 was in a high state (within a few percent of the level of the 
HUT spectrum in Figure 2), so we might expect a somewhat lower ionization 
parameter in the low-state FUSE spectrum, just as we see. 

We find using the N (\nv)/N (\civ) GHRS spectra ratio alone suspect: first,
the two spectral regions were observed six months apart (Crenshaw et al. 1999)
and absorbers have been seen to vary on this timescale, second, \nv\ and 
\civ\ have similar ionization states and the
column density ratio is rather insensitive to the ioinzation parameter 
(small uncertainties in the ratio correspond to large differences in U),
and third, nitrogen abundances have been found to be preferentially enhanced 
in AGN (primarily in high luminosity AGNs, e.g., Hamann \& Ferland 1993), 
and to have significant scatter -- assuming solar abundances
is a large source of uncertainty.  The GHRS N (\nv)/N (\civ) constraint
would seem to be a poor one.  We elected to make our own comparison for
the STIS spectrum alone, which has simultaneous measurements (or interesting
upper limits) for three species.

In order to make a consistent comparison, we used Cloudy with the 
Dumont et al. (1998) spectral energy distribution to model the physical 
conditions of the absorber using the column densities reported by 
Crenshaw \& Kraemer (1999) for the STIS spectrum (in the case of \civ\ 
$\lambda$1549 only an upper limit is available). We 
found a good {\em low} ionization solution, with log U =$-$0.94 and a total 
column density of log N$_H$ = 18.7.  Using instead the Crenshaw \& Kraemer 
(1999) continuum changes the solution to log U = $-$0.38 and  log N$_H$ = 19.3, 
higher ionization and larger column density but still far from 
their high ionization values.  

These low-ionization solutions are not dissimilar from our 
low-ionization solution for the FUSE spectrum.  In fact, the ionization 
parameter (log U = $-$1.7) appears lower for the low-state spectrum as we 
might expect.  Is there a way to decide whether the ``low'' or ``high''
solution is correct?
  
The Chandra spectrum of Kaastra et al. (2000) shows no
\ovii\ and \oviii\ absorption at the velocity of component 1, while
the ``high'' solution would predict easily detectable columns on order 
of log N $\sim 17-18$.  The Chandra spectrum does show \ovi\ absorption
at component 1 velocities, entirely consistent with the FUSE ``low'' solution
and our solution for the STIS column densities.  We conclude that component 1 
is a low column density, low-ionization absorber with the properties 
described above.

\subsection{Components 2 through 5}

Components 2 through 5 correspond to absorption between 
velocities of $-$700 to $-200$\ \kms.  This complex is of special interest
because strong absorption is seen in both longer wavelength UV lines
and in higher ionization X-ray transitions.  

The absorption structure is very similar to that seen in the UV lines
like \civ~$\lambda$1549 (Crenshaw \& Kraemer 1999; Mathur et al. 1999).
Different fitted components, blending, and variability issues conspire to 
make a one-to-one component comparison between the FUSE and HST spectra 
difficult, but the general findings are of interest.  
Crenshaw \& Kraemer (1999) find ionization parameters of a few 
tenths and total hydrogen columns of 10$^{19}$\ to 10$^{20}$\ cm$^{-2}$.  
These conditions produce \ovi\ columns on order of 10$^{15}$\ cm$^{-2}$ and
are consistent with our ``Low'' models of Table 3.

But what of the X-ray absorption?  Kaastra et al. (2000) find
significant column densities for a number of high-ionization species
(e.g.,  log N (\ovii) = 17.2 and log N (\oviii) = 18.1).  
The X-ray absorption appears centered at a velocity of $-$400 \kms\ (using 
z=0.017175), and has a FWHM $\sim$ 600 \kms; Kaastra et al. suggest that that 
large span in velocity is suggestive of contributions of several individual
components of smaller velocity dispersion.  We input the column densities
reported by Kaastra et al. into Cloudy with the Dumont et al. (1998)
continuum and found a good solution similar to that of Kaastra et al.:
log U = 0.43 and log N$_H$ = 21.8.  This model predicts 
log N (\ovi) = 14.8  and log N (\h) = 15.3, less
than what we observe from the sum of the components in the FUSE spectrum,
but enough to account for an individual component or two.  
Such conditions predict very low column densities for \civ\ and \nv.
This may in part explain why we needed to add a component ``2.5'', although 
variability could also be involved.

Based on Cloudy models, an \ovi\ column density on the order of 
log N (\ovi) = 15 would appear to be consistent with {\em both} the X-ray 
and UV absorbers.  We cannot unambiguously say if any individual \ovi\ absorbing
component arises solely from high-ionization or low-ionization gas.  
Simultaneous high-resolution UV-Far-UV-X-ray spectra would help unravel the 
physical conditions of the outflowing material along the line of sight.

\subsection{Component 6}

Component 6 poses a special problem.  The velocity of this component falls
at the apparent peak of the \ovi\ emission, where the covering is most 
uncertain and contaminating molecular H$_2$\ lines are present.  
The formal fitting uncertainties in Table 2 indicate we do not reliably
measure \ovi\ or Lyman $\beta$ absorption in component 6.
Crenshaw \& Kraemer (1999) summarily dismissed this component as uninteresting
given that it was at low velocity and appeared only in Lyman $\alpha$ and 
not in \civ~$\lambda$1549 or \nv~$\lambda$1240. 

\section{Discussion}

Intrinsic absorption is seen in the FUSE spectrum of NGC 5548 in the low-state
in hydrogen Lyman $\beta$ and in \ovi~$\lambda\lambda$1032,1038.
The velocity range and kinematic structure closely resembles that seen
previously by high-resolution HST spectra in the UV 
(Mathur et al. 1999; Crenshaw \& Kraemer 1999).  At least some portion of 
the \ovi-absorbing gas would seem to be identifiable with the previously
studied UV absorber, variability nonwithstanding.

Reanalysis of the STIS data, in conjunction with our FUSE spectrum, indicates
that the high-velocity component 1 is {\em not} high ionization and is
not the X-ray warm absorber.  This reconciles the results of Kaastra et al.
(2000) who report that high-ionization species have a blueshift of 
280$\pm$70 \kms\ (for $z=0.01676$; using $z=0.017175$ the blueshift 
increases by 120 \kms\ to $\sim$400 \kms, consistent with component 3 or 4).
Mathur et al. (1999) suggested that the broad component 3 ($\sim$ 300 \kms
FWHM) could be identified with the X-ray absorber.  The blueshift of
the X-ray absorber and component 3 would tend to support this conclusion,
although there may simply exist material along the line of sight with
a range of physical conditions at these velocities.  
High-resolution, high S/N, simultaneous spectra 
of the UV, far UV, and X-ray regions is required to finally determine the
nature of the absorbing components along the line of sight.

Despite the one-to-one correspondence between the incidence of X-ray
absorption and UV absorption in the 50-60\% of Seyfert galaxies that show
such absorption, there is not one-to-one correspondence
at the level of individual kinematic components.  These outflows are
clearly stratified in some fashion.   The UV absorbers often appear 
as distinct components at a range of velocities, appearing and vanishing
over the time scales of months to years (some of course are more stable).
In Markarian 509 there are seven components, only one of which appears
to be high ionization and it is near zero velocity (Kriss et al. 2000).
In NGC 3783 the X-ray absorber is blueshifted by 440 \kms\ (Kaspi et al. 2000);
the resolution at X-ray wavelengths is not yet sufficient to separate any
kinematic components.  Still, we may conclude that there is no simple
relationship between ionization and velocity, and as yet no clear hint
about their dynamical or physical structure (although radiative acceleration
appears likely in a number of cases including NGC 5548, Srianand 1999).

Thermally driven wind models such as those proposed by
Krolik \& Kriss (1995; 2001) also are consistent with the observations.
These winds are evaporated from the surface of the obscuring torus, and
they have low velocities compatible with those typically seen in Seyfert 1s.
At the critical ionization parameter for evaporation in these models,
there is a broad range of temperatures that can coexist in equilibrium at
nearly constant pressure.  Thus the flow is expected to be strongly
inhomogeneous. High temperature, highly ionized gas causing X-ray absorption
can co-exist with more densely clumped, lower temperature gas that forms
UV absorption lines.

The nature of the narrow emission lines has been only a minor complication for 
studying the intrinsic absorption in far-UV, by coincidence as much as anything
else.  For a high-state spectrum, such as those obtained by HST,
deducing if uncovered narrow emission-lines fill in deep trough bottoms
can have a profound effect on measured columns (e.g., Arav, Korista, \&
de Kool 2001).  The narrow \ovi\ emission lines are of special interest in 
their own right.  How narrow are they?  Are they covered?  

The narrow-line region of NGC 5548 is extended, with deep narrow-band images
showing [\oiii] $\lambda$5007 spanning a region some 4 $\times$ 2.7 kpc
aligned with the weak radio structure (Wilson et al. 1989).
Moore, Cohen, \& Marcy (1996) using optical spectra found that the line widths
and line shifts of narrow emission lines correlated with the critical density
of the transition, rather than ionization potential, and that the low
critical density lines are in fact consistent with the 21 cm redshift
of 0.017175.  The FWHM of these optical narrow lines ranges from
300 to 500 \kms.  This is entirely consistent with what we find for our
preferred uncovered NLR model, and so perhaps we should not be concerned 
that Goad \& Koratkar (1998) found for the low-state FOS spectrum 
(Crenshaw et al.  1993) FWHMs on order of 1000 \kms\ for UV narrow lines like 
\civ~$\lambda$1549.  That was a single epoch, low-resolution spectrum.
Likewise, the NLR models of Kraemer et al. (1998) suggest that 
the high-ionization narrow-line region is rather compact (parsec scale)
and may easily be covered.  However, their model underpredicts the observed
narrow \ovi\ emission lines by nearly an order of magnitude.  

\section{Summary}

We have presented the far-ultraviolet spectrum of NGC 5548 observed by
FUSE in a low state.  Prominent narrow \ovi~$\lambda\lambda$1032,1038
emission lines are present, as are intrinsic absorption lines of 
\ovi\ and hydrogen Lyman $\beta$.  Two rather different assumptions
regarding the narrow \ovi\ emission lines both lead to rather similar
measurements of the absorbing column densities.
Reanalysis of the UV column densities
of the highest velocity component (1) reported by Crenshaw \& Kraemer (1999)
indicates a low total column density and low
ionization parameter and are consistent with our FUSE results and
consistent with the lack of an X-ray absorber at this
velocity (Kaastra et al. 2000).
The absorbers with intermediate velocities of $-200$ to $-700$ \kms
are primarily identified with the same low ionization parameter, low
column density material that absorbers at longer UV wavelengths,
but almost certainly have some non-negligible contribution from 
the high ionization parameter, high column density material 
that results in absorption lines seen in the X-ray.
Still, no single component of \ovi\ absorption can be unambiguously
identified as ``the X-ray absorber.''
High-quality, simultaneous multiwavelength spectra at several
epochs may be required to fully understand the emitting and absorbing 
regions.

\acknowledgments

We thank Nahum Arav, Qirong Yuan, and Steve Kraemer for useful comments.
This work is based on data obtained for the Guaranteed Time Team by the
NASA-CNES-CSA FUSE mission operated by the Johns Hopkins University.
Financial support to U. S. participants has been provided by
NASA contract NAS5-32985.  This research has made use of the NASA/IPAC 
Extragalactic Database (NED) which is operated by the Jet Propulsion 
Laboratory, California Institute of Technology,
under contract with the National Aeronautics and Space Administration.

\begin{deluxetable}{lcccc}
\tabletypesize{\small}
\tablecaption{Model \ovi\ Emission-Line Parameters\tablenotemark{a}}
\tablewidth{0pt}
\tablehead{Feature& Observed $\lambda$ & Flux\tablenotemark{b} & FHWM & Velocity\tablenotemark{c} \\
& (\AA) & ($10^{13}$ ergs s $^{-1}$ cm$^{-2}$) & (\kms) & (\kms) }
\startdata
 & & & & \\
{\em Uncovered NLR Model} & & & & \\
Broad \ovi~$\lambda$1032 & 1048.94 & 2.29$\pm$0.04 & 4841$\pm$89 & $-$200 \\
Broad \ovi~$\lambda$1038 & 1054.72 & 1.14 & 4841 & $-$200  \\
Intermediate \ovi~$\lambda$1032 & 1048.94 & 0.95$\pm$0.04 & 1111$\pm$47 & $-$200  \\ 
Intermediate \ovi~$\lambda$1038 & 1054.72 & 0.47 & 1111 & $-$200  \\
Narrow \ovi~$\lambda$1032 &1049.49$\pm$0.04 & 1.69$\pm$0.10 & 432$\pm$12 & $-$50 \\
Narrow \ovi~$\lambda$1038 & 1055.26 & 1.06$\pm$0.03 & 432 & $-$50  \\
 & & & & \\
{\em Covered NLR Model} & & & & \\
Broad \ovi~$\lambda$1032 & 1048.94$\pm$0.05 & 2.34$\pm$0.05 & 4366$\pm$107 & $-200$  \\
Broad \ovi~$\lambda$1038 & 1054.7 & 1.17 & 4366 & $-200$  \\
Narrow \ovi~$\lambda$1032 &1049.32$\pm$0.05 & 3.27$\pm$0.03 & 658$\pm$9 & $-90$  \\
Narrow \ovi~$\lambda$1038 & 1054.11 & 1.65$\pm$0.03 & 658 & $-90$  \\
\enddata
\tablenotetext{a}{Values without uncertainties are either tied to another parameter or became fixed during the fitting process.}
\tablenotetext{b}{These are observed fluxes.  If the Galactic reddening is assumed to be $E(B-V)=0.02$\ mag for the extinction of Cardelli, Clayton, \& Mathis
(1989), then these fluxes should be increased by 30\%.}
\tablenotetext{c}{Velocities are relative to systemic $z=0.017175$ 
(\h\ 21 cm, NED).}
\end{deluxetable}

\clearpage

\begin{deluxetable}{lccccc}
\tabletypesize{\scriptsize}
\tablecaption{Model Absorption Parameters\tablenotemark{a}}
\tablewidth{0pt}
\tablehead{Feature& Observed $\lambda$ & Column Density & FWHM & Covering\tablenotemark{b} & Velocity\tablenotemark{c} \\
& (\AA) & ($10^{12}$ cm$^{-2}$) & (\kms) & Fraction & (\kms) }
\startdata
{\em Uncovered NLR Model}  & & & & & \\
{\bf \ovi~$\lambda$1032} & & & & & \\
0.5 & 1045.03$\pm$0.11 & 229$\pm$33 & 347$\pm$154 & 1.00 & $-$1320 \\ 
1 & 1045.92$\pm$0.03 & 544$\pm$29 &  278$\pm$17  & 0.99$\pm$0.10 & $-$1070 \\
2 & 1047.19$\pm$0.01 & 296$\pm$57  &  43$\pm$17    & 0.66$\pm$0.05 & $-$700 \\
2.5 & 1047.46$\pm$0.02 & 680$\pm$154 & 62$\pm$11 & 0.41$\pm$0.10 & $-$630 \\
3 & 1047.79$\pm$0.02 & 208$\pm$28  &  142$\pm$12  & 0.75$\pm$0.06 & $-$530  \\
4 & 1048.35$\pm$0.01 & 2121$\pm$242 & 101$\pm$4   & 0.89$\pm$0.02 & $-$370 \\
5 & 1048.86$\pm$0.04 & 793$\pm$68  &  125$\pm$10  & 0.94$\pm$0.08 & $-$230  \\
6 & 1049.78$\pm$0.05 & 25$\pm$18  &  69$\pm$5  & 1.00 & $+$40 \\	
{\bf Ly $\beta$} & & & & \\
0.5 & 1038.75  &    87$\pm$75 &   347 &      1.00  & $-$1320 \\
1 &  1039.63  &     777$\pm$123  & 278   &      0.99 & $-$1070  \\
2 &  1040.89  &     340$\pm$302 &   43  &      0.66  & $-$700 \\
2.5 & 1041.16  &    256$\pm$50  &  62   &      0.41  & $-$630 \\
3 & 1041.49  &     1830$\pm$1406 &   142  &      0.75 & $-$530  \\
4 & 1042.04  &     1589$\pm$600  &  101   &      0.89 & $-$370  \\ 
5 & 1042.55  &      498$\pm$225   &  125   &      0.94  & $-$230 \\
6 & 1043.47  &       77$\pm$32   &  69   &      1.00   & $+$40 \\
 & & &  & \\
{\em Covered NLR Model}  & & & & & \\
{\bf \ovi~$\lambda$1032} & & & & & \\
0.5 & 1045.00$\pm$0.08 & 282$\pm$31 &  289$\pm$35 & 0.99$\pm$0.12 & $-$1330 \\
1 & 1045.89$\pm$0.06 & 473$\pm$104  & 253$\pm$34  & 0.98$\pm$0.06  & $-$1080 \\
2 & 1047.20$\pm$0.01 & 315$\pm$45  &  41$\pm$12  & 0.67$\pm$0.16 & $-$700 \\
2.5 & 1047.45$\pm$0.02 &  688$\pm$139 &  60$\pm$10 & 0.47$\pm$0.04 & $-$630 \\
3 & 1047.72$\pm$0.01 & 313$\pm$38  &  99$\pm$6  &  0.72$\pm$0.03 & $-$550 \\
4 & 1048.30$\pm$0.03 & 1692$\pm$188 &  129$\pm$7 &  0.85$\pm$0.08 & $-$390 \\
5 & 1048.90$\pm$0.01 & 369$\pm$21  &  143$\pm$6  &  0.94$\pm$0.14 & $-$210 \\
6 & 1049.84$\pm$0.11 &  1.1$\pm$1.0  &  78$\pm$36 &  1.00 & $+$50 \\
{\bf Ly $\beta$} & & & & & \\
0.5 & 1038.72 & 125$\pm$220 &  289 &   0.99  & $-$1330 \\
1 & 1039.60 &  816$\pm$262  & 253   &   0.98   & $-$1080 \\
2 & 1040.90 &  325$\pm$102  &  41   &   0.67  & $-$700 \\
2.5 & 1041.16 & 256$\pm$203  &  60  &   0.47  & $-$630 \\
3 & 1041.42 &  1346$\pm$575  &  99 &   0.72   & $-$550 \\
4 & 1042.00 &  2039$\pm$330  &  129 &   0.85  & $-$390 \\
5 & 1042.59 &  560$\pm$145  &  143   &   0.94  & $-$210 \\
6 & 1043.52 &  109$\pm$31  &  78  &   1.00  & $+$50 \\
\enddata
\tablenotetext{a}{Values without uncertainties are either tied to another 
parameter (e.g., the Ly $\beta$ wavelengths) or became fixed during the 
fitting process because of imposed limits.}
\tablenotetext{b}{The covering fraction of the absorbing components is with
respect to all emission components except the narrow emission lines for the
case of the ``uncovered'' model.}
\tablenotetext{c}{Velocities given are the component Gaussian central 
wavelengths relative to systemic $z=0.017175$ (\h\ 21 cm, NED).}
\end{deluxetable}

\clearpage

\begin{deluxetable}{lccccrc}
\tablecaption{Photoionization Models\tablenotemark{a}}
\tablewidth{0pt}
\tablehead{
 & & \multicolumn{2}{c}{Low} & & \multicolumn{2}{c}{High} \\
Component& $N_{\ovi}/N_{\h}$ & log U & log $N_H$ & & log U & log $N_H$ \\
& & & (cm$^{-2}$) & & & (cm$^{-2}$)  }
\startdata
0.5 & 2.63 & $-$1.28 & 18.0 &  & $-$0.16 & 19.5 \\ 
  1 & 0.70 & $-$1.71 & 18.4 &  & 0.26 & 21.1 \\
  2 & 0.87 & $-$1.66 & 18.1 &  & 0.20 & 20.7 \\
2.5 & 2.66 & $-$1.28 & 18.4 &  & $-$0.15 & 20.0 \\
  3 & 0.11 & $-$2.05 & 18.4 &  & 0.56 & 22.0 \\
  4 & 1.33 & $-$1.55 & 18.9 &  & 0.11 & 21.2 \\
  5 & 1.59 & $-$1.48 & 18.5 &  & 0.03 & 20.6 \\
  6 & 0.32 & $-$1.88 & 17.3 &  & 0.42 & 20.4 \\
\enddata
\tablenotetext{a}{Input column densities are for the uncovered NLR model fit.
Given only $N_{\ovi}$ and $N_{\h}$ column densities, both a high (U, $N_H$) and
low (U, $N_H$) solution can be found using Cloudy.}
\end{deluxetable}

\clearpage

\scriptsize

\begin{figure}
\plotone{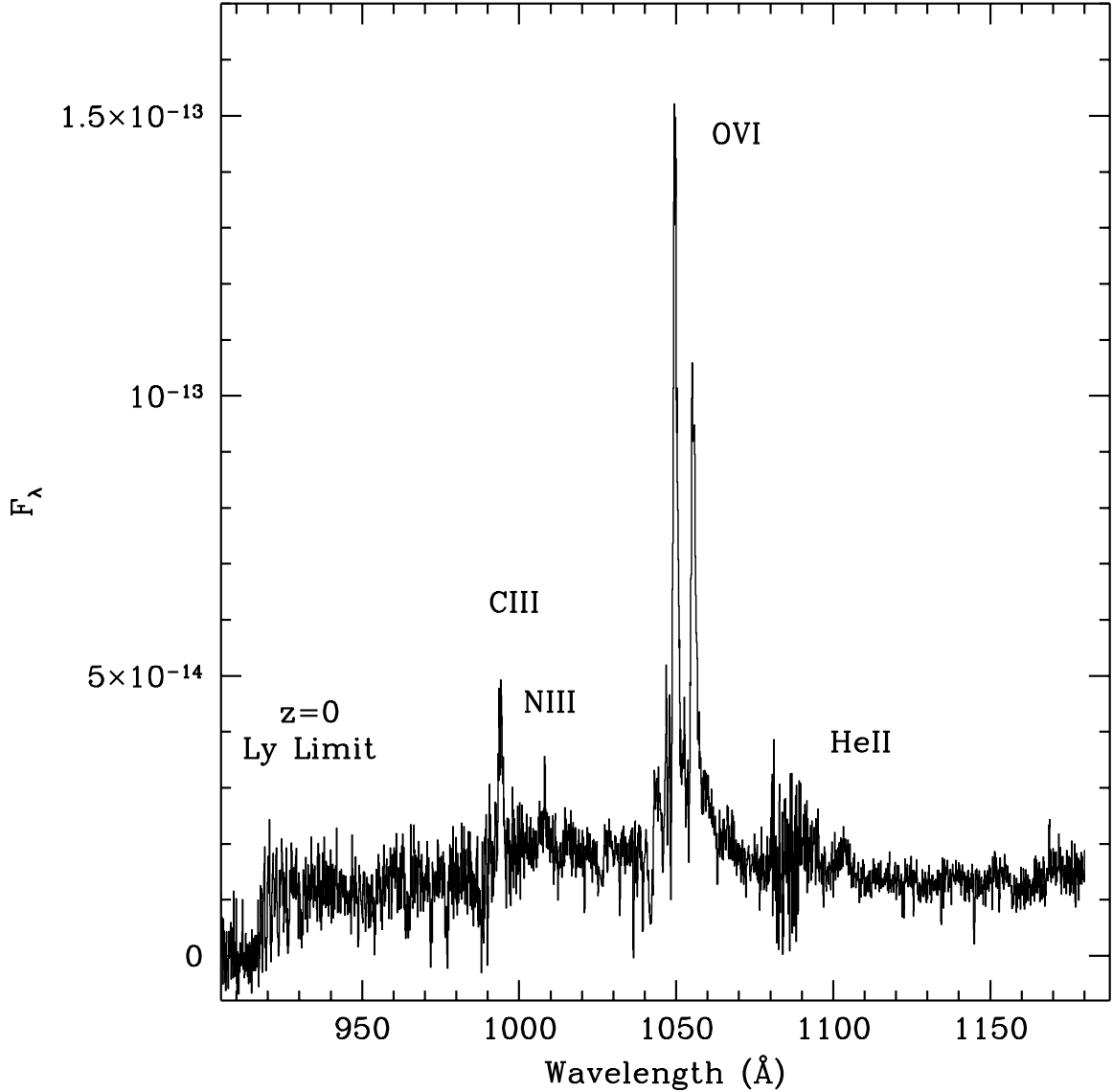}
\caption{
A full-range display spectrum with 0.1 \AA\ bins; this is a coarser binning
than used for analysis and is used here for display only.  
Geocoronal emission lines have been removed.  Prominent features are
labeled.  The flux units are ergs s$^{-1}$ cm$^{-2}$ \AA$^{-1}$.
The low signal-to-noise ratio from 1080-1100 \AA\ results from the 
fact that only a single SiC channel contributes at those wavelengths.}
\end{figure}

\begin{figure}
\plotone{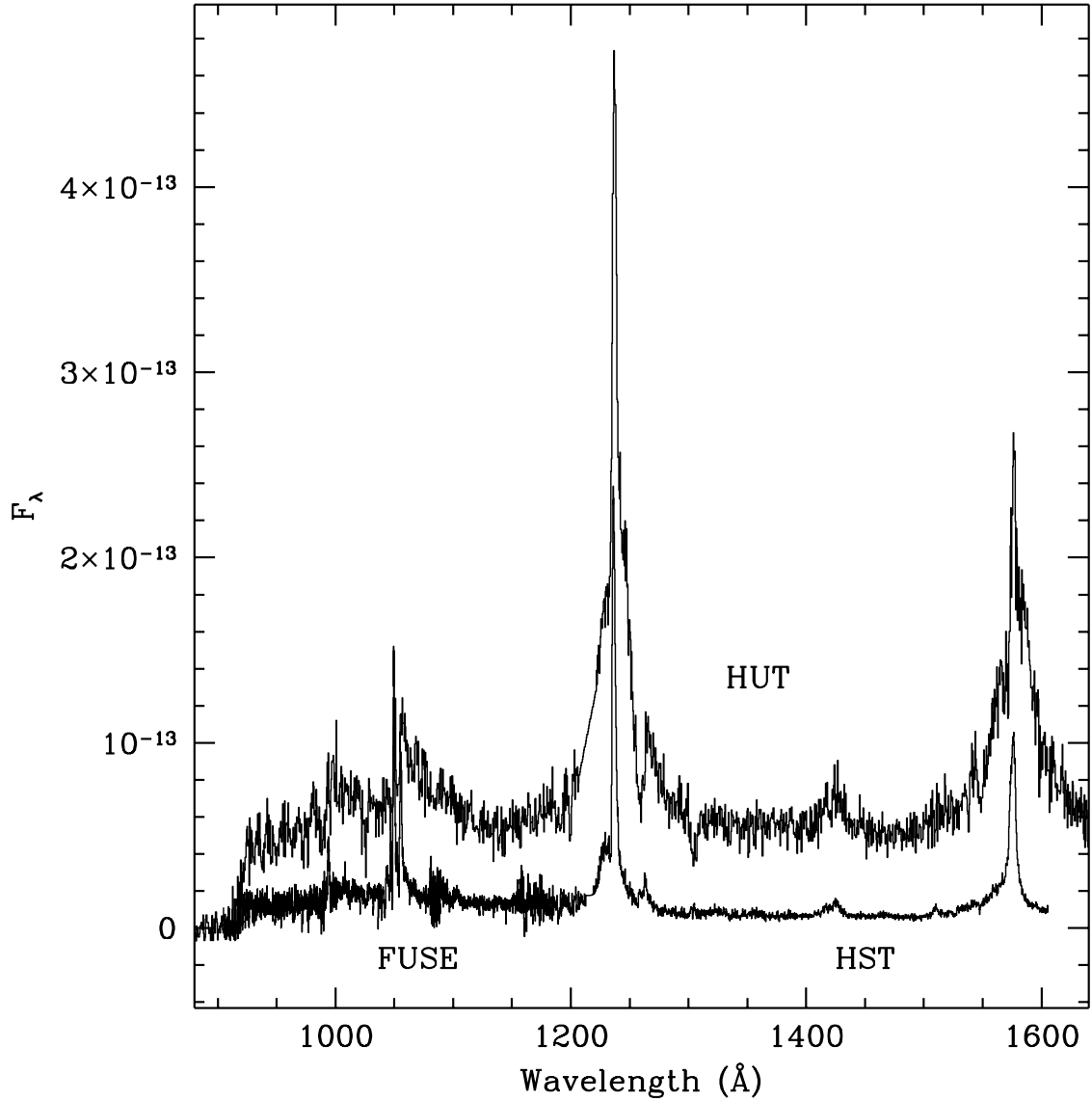}
\caption{
A comparison between the (2000) FUSE spectrum, a previous epoch (1991) low-state
Faint Object Spectrograph (FOS) spectrum from HST (Crenshaw et al. 1993), 
and a Hopkins Ultraviolet Telescope (HUT) spectrum of
NGC 5548 in the high state (Kriss et al. 1997).
}
\end{figure}

\begin{figure}
\plotone{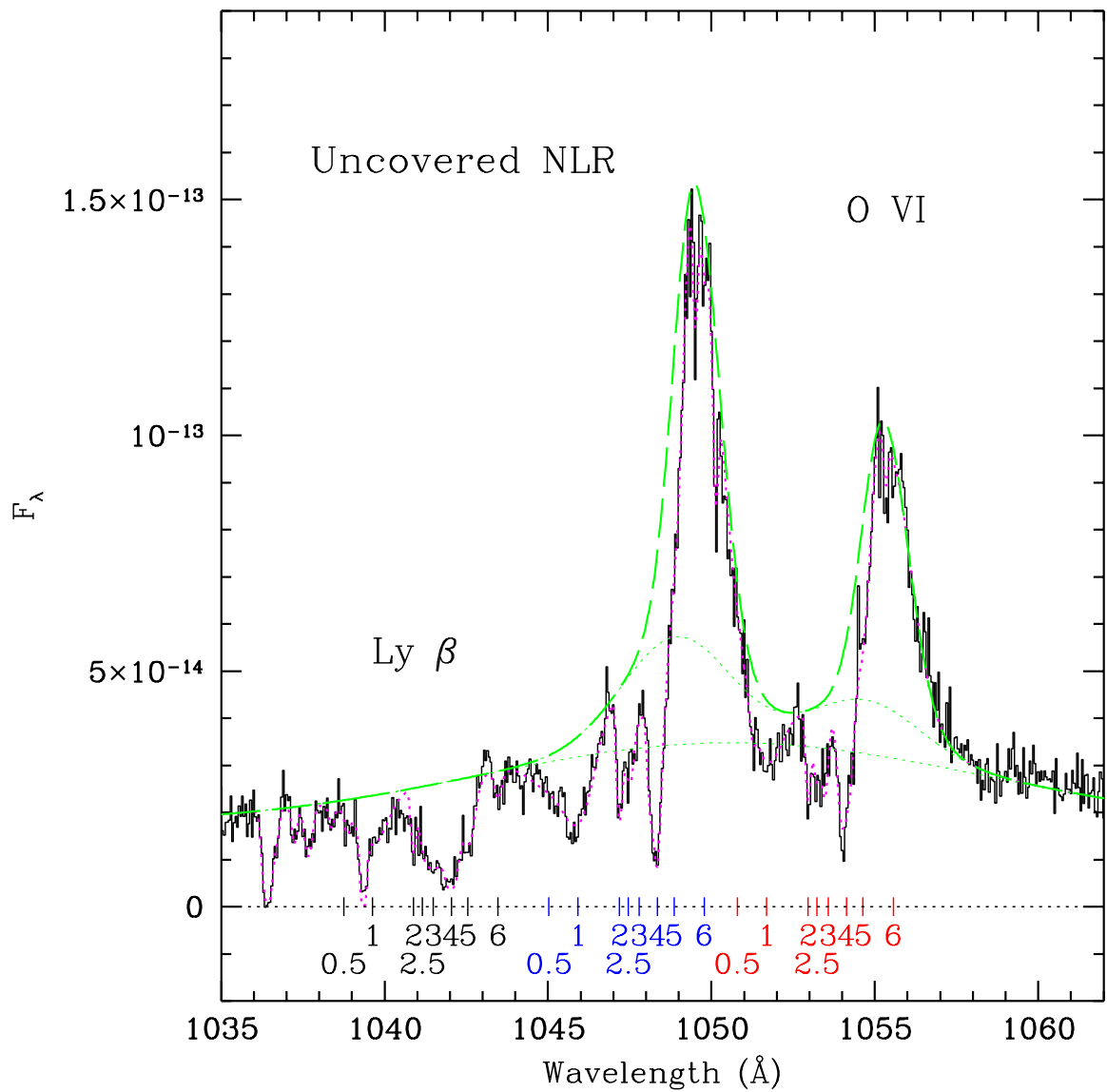}
\caption{ \footnotesize
Our preferred SPECFIT model (dotted magenta line) of the combined-channel
spectrum (0.5 \AA\ bins) of \ovi~$\lambda\lambda$1032,1038
and Ly $\beta$.  The model has $\chi^2 = 1240.2$\ for 801 points and
85 degrees of freedom.  The dashed green line indicates emission components.  
The absorption lines were assumed to at least 
partially cover the broad lines and continuum but not the narrow emission-line 
component.  The dotted line indicates zero flux.  
The central wavelengths of the absorbing components in Table 2 are marked.
}
\end{figure}

\begin{figure}
\plotone{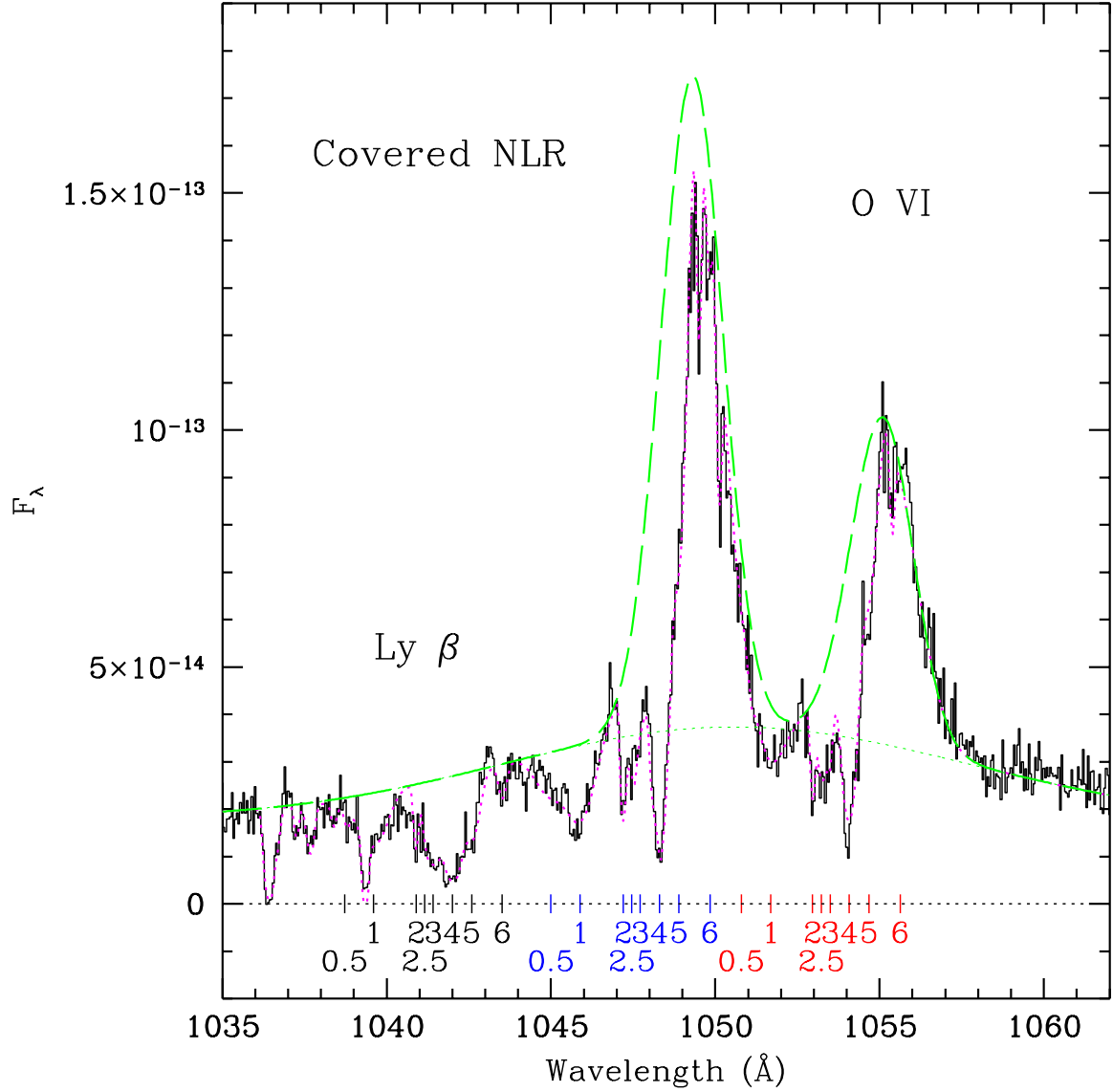}
\caption{ \footnotesize
An alternative SPECFIT model (dotted magenta line) of the combined-channel
spectrum (0.5 \AA\ bins) of \ovi~$\lambda\lambda$1032,1038
and Ly $\beta$.  The model has $\chi^2 = 1356.4$\ for 801 points and
82 degrees of freedom.
The dashed green line indicates emission components.
In this case the absorption lines were assumed to at least partially cover
all emission components.  The dotted line indicates zero flux.  
The central wavelengths of the absorbing components in Table 2 are marked.
}
\end{figure}

\begin{figure}
\plotone{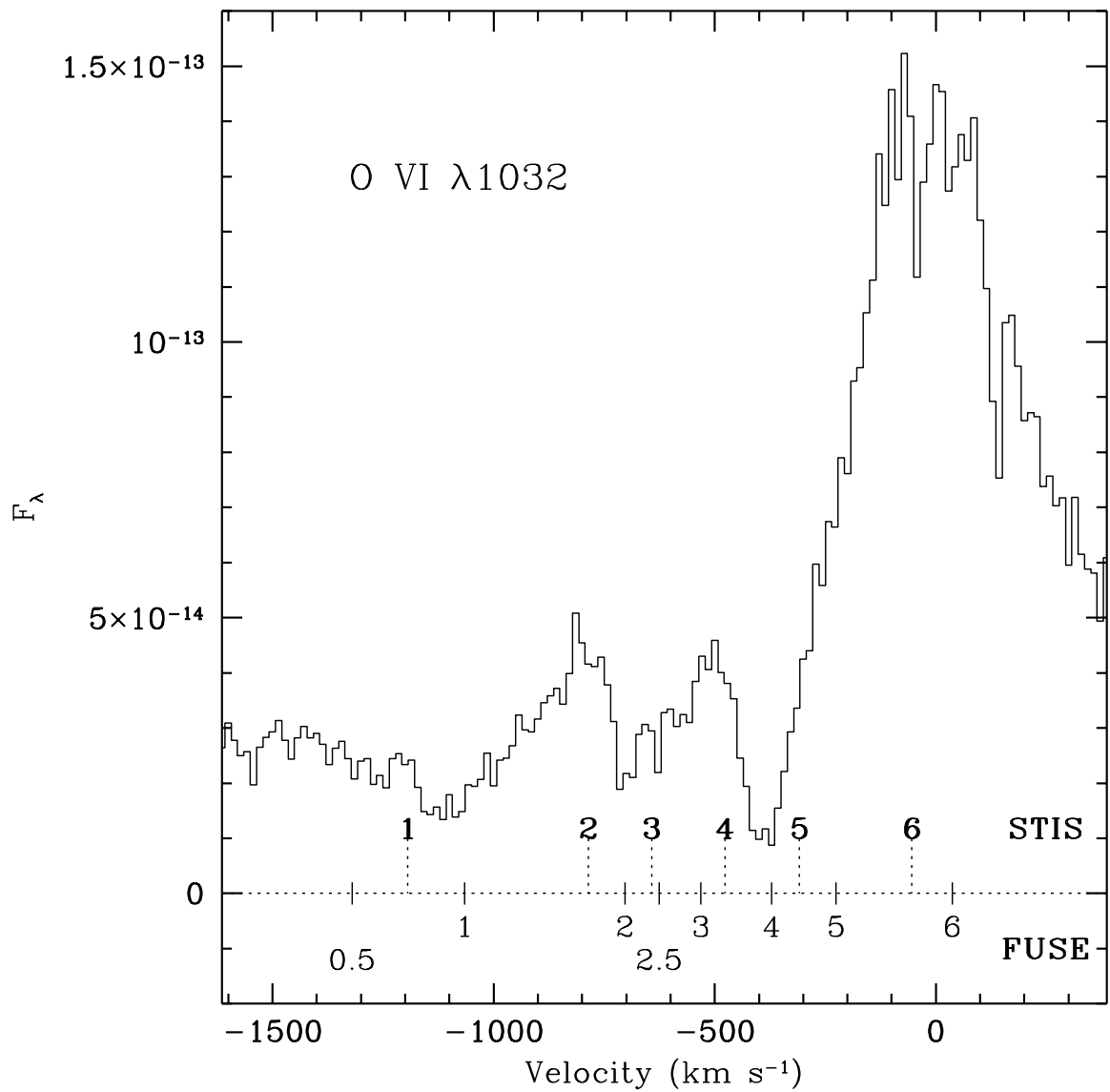}
\caption{ \footnotesize
Blow up of the blue doublet of \ovi\ on a velocity scale 
(assuming $z=0.017175$).  Numbers/tickmarks below zero show the positions
of the fitting components just as in the uncovered NLR model of Figure 3.
Numbers/tickmarks above zero indicate the velocities of Crenshaw \& Kraemer's 
(1999) components 1-6 based on the HST STIS spectrum. It would appear 
that there is a velocity shift $\sim100$ \kms\ or less between the two,
probably the result of calibration uncertainties.
}
\end{figure}

\end{document}